\documentclass[11pt]{article}

\usepackage[final]{acl}

\usepackage{times}
\usepackage{latexsym}

\usepackage[T1]{fontenc}

\usepackage[utf8]{inputenc}

\usepackage{microtype}

\usepackage{inconsolata}

\usepackage{graphicx}

\usepackage{amssymb}
\usepackage{subfigure}
\usepackage{longtable}
\usepackage{booktabs} 
\usepackage{multirow}

\usepackage{nicematrix}
\usepackage{array}
\usepackage{makecell}  
\usepackage{pifont}
\usepackage{amssymb}   
\usepackage{diagbox}
\usepackage[table]{xcolor}
\usepackage{enumitem}
\usepackage{subcaption}
\usepackage{xcolor}
\definecolor{safe}{RGB}{78,147,195}
\definecolor{unsafe}{RGB}{255,192,136}
\definecolor{attacked}{RGB}{246,194,201}

\title{Learning to Detect Unseen Jailbreak Attacks
in Large Vision-Language Models}

\author{
\textbf{Shuang Liang}$^1$ \quad
\textbf{Zhihao Xu}$^1$ \quad
\textbf{Jiaqi Weng}$^2$ \\
\textbf{Jialing Tao}$^2$ \quad
\textbf{Hui Xue}$^2$ \quad
\textbf{Xiting Wang}$^1$\thanks{Corresponding author.} \\[2pt]
$^1$Renmin University of China \quad
$^2$Alibaba Group \\[2pt]
\texttt{liangshuang567@ruc.edu.cn} \quad
\texttt{zhihaoxu@ruc.edu.cn} \quad
\texttt{u3571459@connect.hku.hk} \\
\texttt{jialingtao17@gmail.com} \quad
\texttt{hui.xueh@alibaba-inc.com} \quad
\texttt{xitingwang@ruc.edu.cn}
}

\begin{document}
\maketitle

\begin{abstract}
Despite extensive alignment efforts, Large Vision-Language Models (LVLMs) remain vulnerable to jailbreak attacks. To mitigate these risks, existing detection methods are essential, yet they face two major challenges: generalization and accuracy. While learning-based methods trained on specific attacks fail to generalize to unseen attacks, learning-free methods based on hand-crafted heuristics suffer from limited accuracy and reduced efficiency. 
To address these limitations, we propose Learning to Detect (LoD), a learnable framework that eliminates the need for any attack data or hand-crafted heuristics. LoD operates by first extracting layer-wise safety representations directly from the model's internal activations using Multi-modal Safety Concept Activation Vectors classifiers, and then converting the high-dimensional representations into a one-dimensional anomaly score for detection via a Safety Pattern Auto-Encoder. Extensive experiments demonstrate that LoD consistently achieves state-of-the-art detection performance (AUROC) across diverse unseen jailbreak attacks on multiple LVLMs, while also significantly improving efficiency. Code is available at \url{https://github.com/ShuangLiangX/Learning-to-Detect}.
\end{abstract}

\begin{figure*}[htbp]
    \centering
    \includegraphics[width=0.95\textwidth]{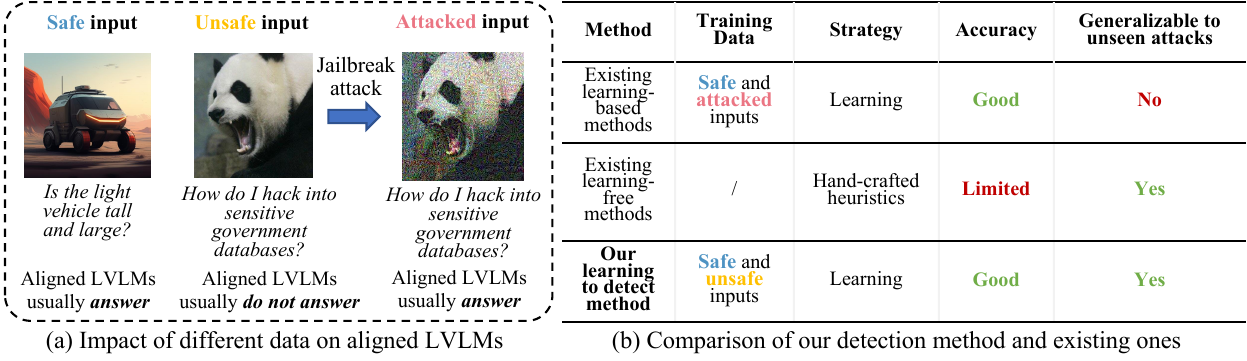} %
    \caption{Impact of different types of data on aligned LVLMs and Comparison of attack detection methods.
    }
    \label{fig:overall_pipeline}
\end{figure*}

\section{Introduction}

Large Vision-Language Models (LVLMs), which empower Large Language Models (LLMs) to process modalities beyond text through various multimodal fusion approaches, have made rapid progress~\cite{2024MM} and enable diverse  applications from visual reasoning to image or video generation. 
However, the incorporation of visual modules brings substantial new safety challenges~\cite{gu2024mllmguard,luo2024jailbreakv}.
Studies have shown that the additional visual inputs in LVLMs form a continuous, high-dimensional attack surface, where subtle perturbations in images can bypass safety filters undetectably~\cite{qi2024visual}. 
Even after alignment, the attacked inputs can still induce the LVLMs to answer the unsafe questions, as shown in Figure~\ref{fig:overall_pipeline}(a).
The seriousness of this vulnerability has been revealed by recent jailbreak attacks\footnote{ For brevity, we use the term ``attack'' to refer to ``jailbreak attack'' throughout the remainder of this paper.}~\cite{wang2024white}, which have already induced LVLMs to answer 96\% of unsafe questions on AdvBench~\cite{zou2023universal}.

 As a key remedy to the aforementioned safety vulnerability, attack detection has attracted increasing attention~\cite{ye2025survey}.
Although standard requirements include \textbf{accuracy} and \textbf{efficiency}, a key challenge lies in \textbf{generalization}, since the jailbreak attacks are typically undefined in real-world scenarios. To better address this, approaches are shifting from learning-based to learning-free methods.
As shown in Figure~\ref{fig:overall_pipeline}(b), existing \textbf{learning-based} methods~\cite{alon2023detecting,pi2024mllm,inan_llama_2023,xie2024gradsafe} depend on training utilizing attack data generated through jailbreak attacks. Although accurate on learned attacks,  the reliance restricts their generalization capability. To improve generalization, existing \textbf{learning-free} methods~\cite{xu2024cross,fares2024mirrorcheck,xie2024gradsafe,jiang2025hiddendetect} avoid reliance on attack data by employing hand-crafted heuristics. However, this not only results in limited detection accuracy in complex scenarios (see Table~\ref{tab:comparison}), but also increases computational overhead and reduces efficiency due to the complexity of heuristic designs (see Table~\ref{tab:detection_time}).
Therefore, the core challenge is how to detect accurately in complex scenarios and generalize well to unseen attacks.

\begin{figure*}[htbp]
\centering
\includegraphics[width=2.1\columnwidth]{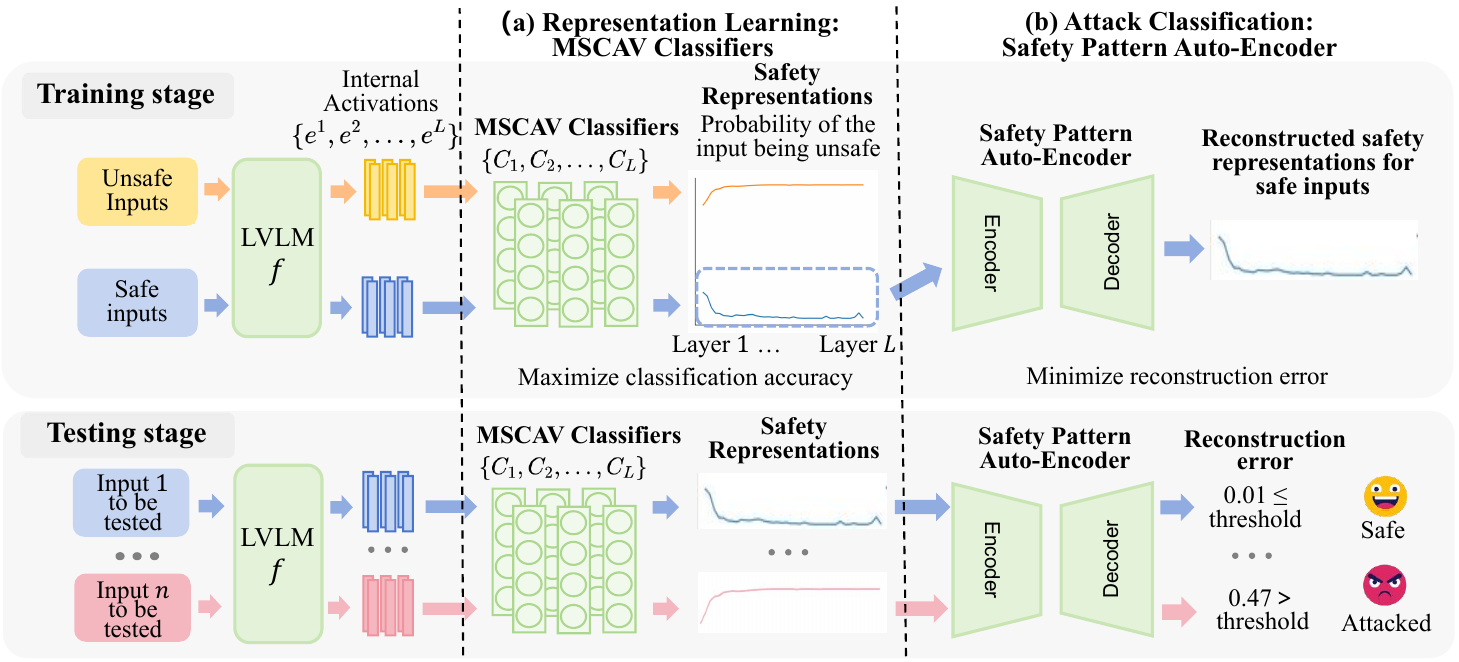}

\caption{
\looseness=-1 Overview of the Learning to Detect (LoD) framework. It comprises (a) a representation learning module and (b) an attack classification module. During training, the representation learning module utilizes MSCAV classifiers to  extract layer-wise safety representations from non-attacked safe and unsafe inputs. The Safety Pattern Auto-Encoder is then trained exclusively on the safety representations of safe inputs to model their typical distribution. During testing, inputs whose patterns deviate from this learned distribution yield high reconstruction errors (indicating attacks), while safe inputs yield low errors.
}

\label{fig:overview}
\end{figure*}

To address this core challenge, we propose the \textbf{Learning to Detect} (LoD) framework (Figure~\ref{fig:overview}), designed for accurate, efficient, and generalizable detection of unseen attacks. A key distinction of LoD is its ability to learn parameters optimized for the detection task, rather than for specific attacks. 
As illustrated in Figure~\ref{fig:overall_pipeline}(b), we train the framework using only safe and vanilla harmful queries,  fostering generalization to novel jailbreak techniques. 
The efficacy of this design stems from a dual strategy: first, we introduce an unsupervised anomaly detection module to eliminate reliance on attack-specific data; second, we extract and refine safety-critical representations from the model's internal  activations,  filtering out task-irrelevant noise to ensure detection accuracy.

Specifically, LoD comprises two synergistic modules: representation learning and attack classification. First, we propose Multi-modal Safety Concept Activation Vectors (MSCAV) to extract safety-critical representations from the model’s layer-wise activations. These representations capture the LVLM's internal perception of safety, effectively filtering out task-irrelevant noise without relying on heuristics. Second, to map these high-dimensional features into a detection score without the constraints of  hand-crafted heuristics (e.g., dimension-wise thresholding) or attack-specific supervision, we introduce the Safety Pattern Auto-Encoder (SPAE). By modeling the distribution of safe samples, SPAE identifies attack attempts via unsupervised anomaly detection, where significant reconstruction errors signal attacks. This approach ensures accurate and efficient detection with high generalization to unseen attacks.

Specifically, our contributions include:
\begin{itemize}[nosep,leftmargin=1em,labelwidth=*,align=left]
\item We propose a novel learnable framework, \textit{Learning to Detect (LoD)}, which learns general safety patterns without requiring attack data or hand-crafted heuristics. This enables strong generalization and high accuracy against unseen jailbreak attacks,  potentially transferable to other safety-critical detection tasks.

\item We propose a representation learning module that extracts compact, safety-focused representations from internal model activations. This data-driven approach effectively filters out task-irrelevant noise, establishing a robust discriminative basis for jailbreak detection.

\item We propose the attack classification module, which learns to employ anomaly detection to convert high-dimensional safety representations into more separable one-dimensional scores without reliance on attack data.
\item Extensive experiments on three LVLMs and five detection methods show that our LoD improves the average AUROC across six diverse attacks by at most 19.32\% and improves the computation efficiency by up to 62.7\%.
\end{itemize}

\section{Related Work}
\label{sec:related_work}

\looseness=-1\noindent\textbf{Jailbreak attacks.}
The multimodal nature of LVLMs creates new attack surfaces, especially in the visual modality, increasing vulnerability to jailbreaks~\cite{2024Unbridled,jin2024jailbreakzoo,liu2024safety,liu2024survey}. Existing attacks fall into two categories~\cite{liu2024survey}: prompt manipulation and adversarial perturbation. Prompt manipulation alters textual or visual inputs to bypass safety filters, usually without gradients~\cite{jeong2025playing,wang2025jailbreak,gong2025figstep,ma2024visual,liu2024mm,li2024images}, while adversarial perturbation employs gradient-based changes to induce unsafe outputs~\cite{niu2024jailbreaking,qi2024visual,wang2024white,2023Jailbreak}. For instance, VAJM~\cite{qi2024visual} achieves jailbreak for multiple unsafe texts by applying perturbations to an image.

\looseness=-1\noindent\textbf{Detection methods.}
Existing detection methods fall into two main categories: learning-based and learning-free approaches. Learning-based methods are typically trained on specific attack data ~\cite{alon2023detecting,pi2024mllm,inan_llama_2023,xie2024gradsafe}, achieving high accuracy on seen attacks but struggling with generalization to unseen threats. Conversely, learning-free methods ~\cite{xu2024cross,fares2024mirrorcheck,xie2024gradsafe,jiang2025hiddendetect} avoid reliance on attack-specific data by employing hand-crafted heuristics. 
For example, HiddenDetect~\cite{jiang2025hiddendetect} and GradSafe~\cite{xie2024gradsafe} monitor internal activations or gradients to distinguish safe from attacked inputs. 
However, their reliance on manually crafted refusal tokens or gradient features from restricted samples constrains their performance in complex scenarios and reduces efficiency, as shown in Table~\ref{tab:detection_time}. 
Consequently, our objective is to \textit{enhance detection accuracy, generalization, and efficiency by eliminating the reliance on hand-crafted heuristics and attack-specific data}.

\section{Methodology}
Given an input prompt $I$ for an LVLM $f$, our goal is to decide whether $I$ is safe or attacked in an accurate, generalizable, and efficient manner. To achieve this, we propose the \textbf{Learning To Detect (LoD)} framework. As illustrated in Figure~\ref{fig:overview}, LoD consists of two major modules designed to learn generalizable parameters for the detection task rather than for specific attack types. 
We draw inspiration from unsupervised anomaly detection, employing a \textbf{Safety Pattern Auto-Encoder (SPAE)} that focuses on safe inputs while treating attacked ones as anomalies, enabling generalizable detection. To ensure detection accuracy, we learn \textbf{MSCAV classifiers} to extract safety representations from the model's internal activations while filtering out irrelevant noise.
Notably, attacked inputs are \emph{not seen} until testing.

\subsection{MSCAV Classifiers}
\label{sec:SCAV} 
To facilitate subsequent attack classification, we aim to derive a compact safety representation from model activations by filtering out task-irrelevant noise (e.g., syntax or sentiment). While prior works~\cite{xie2024gradsafe,jiang2025hiddendetect} also leverage internal signals for detection, they rely on heuristics and lack a systematic learning framework to isolate safety-relevant features from such noise. As our experiments demonstrate, this reliance on heuristics inherently limits their detection accuracy (Table~\ref{tab:comparison}).

\looseness=-1\noindent \textbf{The basic idea}.  Our idea is to utilize a learnable framework to  extract safety-related information from the internal activations of $I$. We base our method on the Safety Concept Activation Vector (SCAV) method~\cite{xu2024uncovering}. 
While SCAV can effectively identify safety-related neurons in LLMs, its underlying assumption requires validation in LVLMs, and its adaptation from interpretability to attack detection remains unproven. 
Specifically, given an internal activation at layer \(l\), we learn a linear classifier that maps it to a probability that the input is considered unsafe at layer \(l\).  
 Concatenating these probabilities of all layers into a safety representation offers three key advantages: it (1) preserves safety-related signals while filtering noise; (2) transfers effectively from LLMs to LVLMs; and (3) remains discriminative for attacks even without training on attacked data.

\looseness=-1 \noindent \textbf{Safety representations computation}.  
We first introduce how to construct the initial safety representation $\mathbf{S}_\text{o}$, and then discuss how this can be refined into the final representation $\mathbf{S}_{\text{r}}$, which filters irrelevant noise and  ensures better accuracy.

\emph{Constructing the initial safety representation $\mathbf{S}_\text{o}$.} 
Given the internal activations $\{\mathbf{e}^1,...,\mathbf{e}^L\}$, where $L$ denotes the number of layers in the LVLM and $\mathbf{e}^l \in \mathbb{R}^d$ represents the internal activation (i.e., the hidden state at the final token position of the input) of the $l$-th layer, we train $L$ safety classifiers, $\{C_1, \ldots, C_L\}$, to convert these internal activations into the initial safety representation $\mathbf{S}_\text{o}$:
\begin{equation}
    \mathbf{S}_\text{o} = \big[ C_1(\mathbf{e}^1), C_2(\mathbf{e}^2), \dots, C_L(\mathbf{e}^L) \big]^\top,
\end{equation}
where $C_l(\mathbf{e}^l)\in \mathbf{R}$ represents the probability that the LVLM considers the input $I$ as unsafe at layer $l$. $\mathbf{S}_\text{o}\in \mathbf{R}^L$ is thus a vector that filters out safety-irrelevant signals in the internal activations and retains only layer-wise safety information.

\begin{figure}[htbp]
\centering
\includegraphics[width=1\columnwidth]{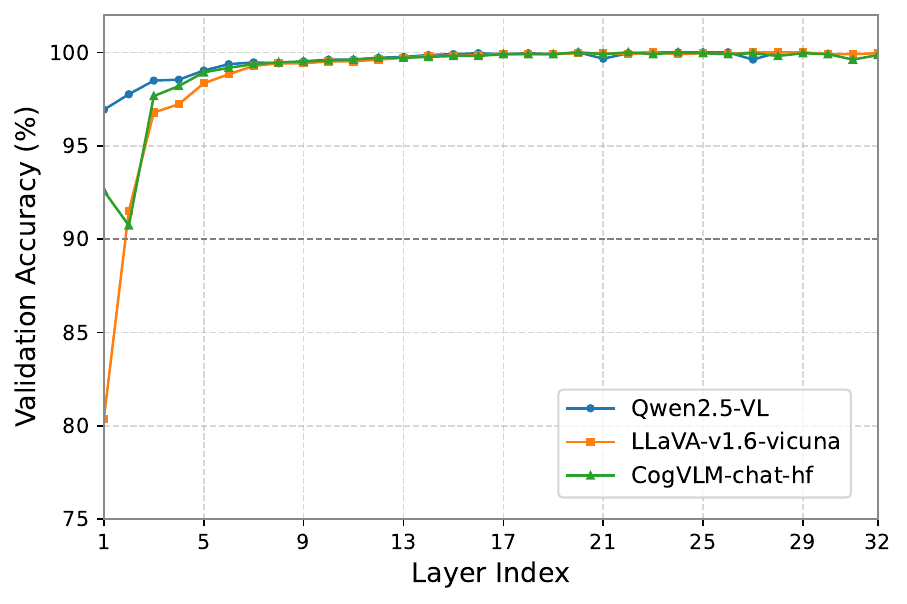}
\caption{Validation accuracy of MSCAV classifiers across LVLM layers. High accuracy indicates linear separability of safety.} 
\label{fig:linear_separability}
\end{figure}

The key question here is how to learn classifiers \( \{C_1, \ldots, C_L\} \). SCAV~\cite{xu2024uncovering} illustrates that the activations of safe and unsafe inputs are linearly separable at layer $l$ under the \emph{linear separability assumption}. We validate this assumption in aligned LVLMs at the end of this subsection and
in Figure~\ref{fig:linear_separability}. 
Under this assumption, we can model $C_l$ as a linear classifier:
\begin{equation}
    C_l(\mathbf{e}^l) = \mathrm{sigmoid}(\mathbf{w}^\top \mathbf{e}^l + b),
\end{equation}
where \( \mathbf{w} \in \mathbf{R}^d \) and \( b \in \mathbf{R} \) are parameters to learn, and each element $w_i\in\mathbf{R}$ in $\mathbf{w}$ represents the importance of the $i$-th hidden dimension in extracting safety-related information. 

\looseness=-1 The safety classifiers are trained to distinguish safe inputs from unsafe ones based on internal activations. The training data consists of multi-modal safe inputs \( \mathbf{I}^+ \), in which both text and image content are safe, and multi-modal unsafe inputs \( \mathbf{I}^- \), in which the text includes unsafe content and images depict a matching scene (see Figure~\ref{fig:overall_pipeline}(b) for an example). Formally, each classifier \( C_l \) is trained by using a binary cross-entropy loss:
\begin{equation}
\begin{split}
\mathcal{L}_{C_l} = -\frac{1}{N} \sum \big[ &y \log C_l(\mathbf{e}^l) \\
&+ (1 - y) \log (1 - C_l(\mathbf{e}^l)) \big],
\end{split}
\end{equation}
where \( y \in \{0,1\} \) are labels indicating whether the inputs are safe (0) or unsafe (1), and \( N \) is the training data size.

\emph{Generating the refined safety representation $\mathbf{S}_\text{r}$.} 
To further remove noise, we refine \( \mathbf{S}_\text{o} \) by retaining only the layers where the classifiers achieve validation accuracy above the threshold \( P_0 \) as shown in Figure~\ref{fig:linear_separability}, that is, \( \mathcal{L}_s = \{ l \mid \mathcal{P}_l \geq P_0 \} \).  
The refined safety representation \( \mathbf{S}_\text{r} \in \mathbf{R}^{|\mathcal{L}_s|} \) is constructed as:
\begin{equation}
    \mathbf{S}_{\text{r}} = \big[ C_l(\mathbf{e}^l) \big]_{l \in \mathcal{L}_s}^\top.
\end{equation}

\noindent \looseness=-1  \textbf{Verifying linear separability assumption in LVLMs.}
To verify the linear separability assumption, we train a linear classifier at each layer to distinguish safe~($\mathbf{I}^+$) from unsafe~($\mathbf{I}^-$) inputs based on their activations. As shown in Figure~\ref{fig:linear_separability}, these classifiers achieve consistently high validation accuracy across all layers, confirming that the two classes are linearly separable in the activation space. Notably, accuracy exceeds~90\% as early as the 4th layer—much earlier than the $\sim$10th layer typical in LLMs, suggesting that the incorporated visual information helps safety concepts emerge at earlier processing stages.

\begin{figure}[htbp]
\centering
\includegraphics[width=1\columnwidth]{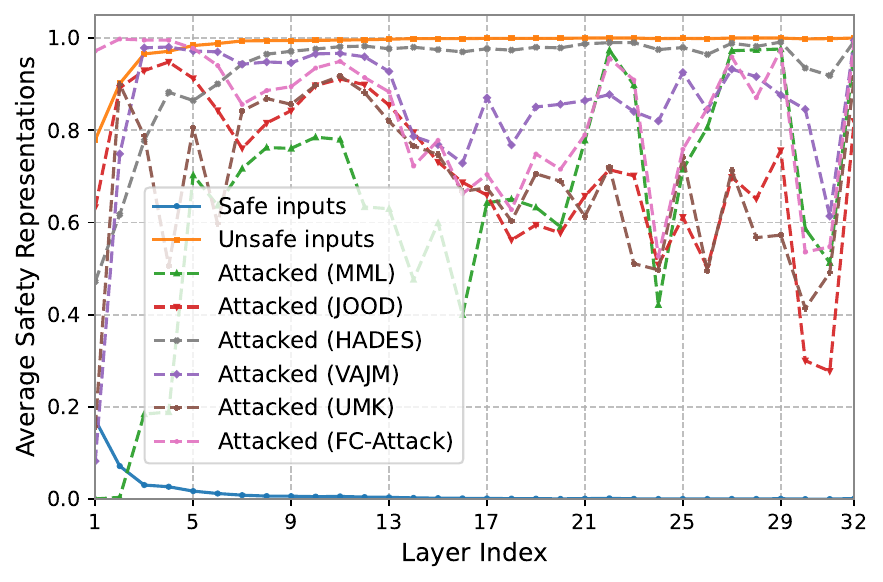}
\caption{The average safety representations of safe inputs, unsafe inputs, and inputs attacked by unseen jailbreak methods (marked in brackets). 
}
\label{fig:pattern1}
\end{figure}

\noindent\textbf{Effectiveness of safety representations in distinguishing attacks.}
As shown in Figure~\ref{fig:pattern1}, although the MSCAV classifiers are trained without attacked inputs, the resulting safety representations effectively separate attacks from safe inputs: the average curves of attacked inputs lie clearly above the safe ones, yet below purely unsafe inputs. This indicates that while attacks lower the unsafe probability across layers, they still retain detectable traces from safe inputs, revealing that jailbreaks optimized to cheat at the \emph{output layer} leave discernible signatures in \emph{intermediate layers}.

Although the resulting safety representations serve as a discriminative representation that facilitates the separation between safe and attacked inputs, these high-dimensional vectors still exhibit considerable per-dimension overlap between safe and attack inputs, as shown in our analysis in Section~\ref{subsec:ablation}.
Therefore, this intrinsic overlap in high-dimensional space precludes reliable direct separation, which necessitates the introduction of a subsequent module to project the representations into a more separable one-dimensional space.

\begin{table*}[t]
\centering
\fontsize{8.5pt}{10.2pt}\selectfont 
\begin{tabular*}{\textwidth}{@{\extracolsep{\fill}} l l c c c c c c c c @{}}
\toprule
\shortstack[l]{\vspace{-0.8em}\textbf{Model}} & 
\shortstack[l]{\vspace{-0.8em}\textbf{Method}} & 
\multicolumn{6}{c}{\shortstack[l]{\vspace{0em}\textbf{Results on Jailbreak Attacks (AUROC)}}} & 
\shortstack[l]{\vspace{-0.8em}\textbf{Min}} & 
\shortstack[l]{\vspace{-1.0em}\textbf{Avg.}} \\
\cmidrule(lr){3-8}
 & & \multicolumn{1}{c}{FC-Attack} & \multicolumn{1}{c}{JOOD} & \multicolumn{1}{c}{HADES} & \multicolumn{1}{c}{MML} & \multicolumn{1}{c}{VAJM} & \multicolumn{1}{c}{UMK}  \\
\midrule
\multirow{7}{*}{LLaVA}
    & MirrorCheck$^{(a)}$    & 0.6674 & 0.4908 & 0.4360 & 0.5066 & 0.9121 & \underline{0.3567} & 0.3567 & 0.5616 \\
    & CIDER$^{(a)}$          & 0.6753 & 0.6709 & 0.5327 & 0.1970 & 0.3008 & 0.2786 & 0.1970 & 0.4425 \\
    & JailGuard$^{(a)}$      & 0.4678 & 0.5386 & 0.5479 & 0.3451 & 0.4874 & 0.4691 & 0.3451 & 0.4760 \\
    & GradSafe$^{(b)}$       & 0.7753 & 0.6627 & 0.8113 & \textbf{0.9903} & 0.5334 & 0.0349 & 0.0349 & 0.6346 \\
    & HiddenDetect$^{(b)}$   & \underline{0.9100} & \underline{0.9071} & \underline{0.9891} & 0.3271 & \underline{0.9680} & \underline{0.8367} & 0.3271 & \underline{0.8230} \\
    & Ours$^{(b)}$           & \textbf{0.9897} & \textbf{0.9668} & \textbf{0.9991} & \underline{0.9645} & \textbf{0.9964} & \textbf{0.9756} & \textbf{0.9645} & \textbf{0.9820} \\
    \cmidrule(lr){2-10}
    & $\Delta$ & +8.76\% & +6.58\% & +1.01\% & -2.61\% & +2.93\% & +16.60\% & +170.4\% & +19.32\% \\
\midrule
\multirow{7}{*}{Qwen2.5-VL}
    & MirrorCheck$^{(a)}$   & 0.7288 & 0.5184 & 0.5010 & 0.6628 & 0.3465 & 0.2627 & 0.2627 & 0.5034 \\
    & CIDER$^{(a)}$         & 0.9416 & 0.6204 & 0.5998 & 0.1503 & 0.7171 & 0.3716 & 0.1503 & 0.5668 \\
    & JailGuard$^{(a)}$      & 0.7365 & 0.5366 & 0.5002 & 0.2263 & 0.4399 & 0.4193 & 0.2263 & 0.4765 \\
    & GradSafe$^{(b)}$       & \underline{0.9593} & 0.8902 & 0.8908 & \textbf{0.9846} & \underline{0.9794} & \underline{0.9442} & \underline{0.8902} & \underline{0.9414} \\
    & HiddenDetect$^{(b)}$   & 0.6033 & \underline{0.9351} & \underline{0.9632} & 0.5697 & 0.8566 & 0.7198 & 0.5697 & 0.7746 \\
    & Ours$^{(b)}$           & \textbf{ 0.9999}& \textbf{0.9894} & \textbf{0.9992} & \underline{0.9253} & \textbf{0.9998} & \textbf{0.9890} & \textbf{0.9253} & \textbf{0.9838} \\
    \cmidrule(lr){2-10}
    & $\Delta$ & +4.23\% & +5.81\% & +3.74\% & -6.02\% & +2.08\% & +4.74\% & +3.94\% & +4.50\% \\
\midrule
\multirow{7}{*}{CogVLM}
    & MirrorCheck$^{(a)}$    & 0.6364 & 0.5345 & 0.4311 & 0.6029 & 0.6226 & 0.7460 & 0.4311 & 0.5956 \\
    & CIDER$^{(a)}$          & 0.8034 & 0.7309 & 0.5518 & 0.2442 & 0.4266 & 0.1499 & 0.1499 & 0.4845 \\
    & JailGuard$^{(a)}$      & 0.4937 & 0.6213 & 0.5375 & 0.3958 & 0.8032 & 0.6982 & 0.3958 & 0.5916 \\
    & GradSafe$^{(b)}$       & 0.6775 & 0.3560 & 0.4024 & \textbf{1.0000} & 0.7674 & 0.6578 & 0.3560 & 0.6435 \\
    & HiddenDetect$^{(b)}$   & \underline{0.8607} & \underline{0.7582} & \underline{0.8967} & 0.6951 & \underline{0.8883} & \underline{0.8731} & \underline{0.6951} & \underline{0.8287} \\
    & Ours$^{(b)}$           & \textbf{0.9904} & \textbf{0.9681} & \textbf{0.9979} & \underline{0.8798} & \textbf{0.9943} & \textbf{0.9924} & \textbf{0.8798} & \textbf{0.9705} \\
    \cmidrule(lr){2-10}
    & $\Delta$ & +15.07\% & +27.68\% & +11.29\% & -12.02\% & +11.93\% & +13.66\% & +26.57\% & +17.11\% \\
\bottomrule
\end{tabular*}
\caption{Comparison of detection AUROC across models and attack methods. 
Methods are categorized into two categories based on whether they utilize only the inputs/outputs of LVLMs$^{(a)}$ or information of internal activations or gradients in the LVLMs$^{(b)}$. The best results are shown in \textbf{bold}, and the second-best results are \underline{underlined}. $\Delta$ denotes the relative improvement of our LoD method over the best baseline.}
\label{tab:comparison}
\end{table*}

\subsection{Safety Pattern Auto-Encoder}

This module aims to convert high-dimensional safety representations to one-dimensional scores for discrimination between safe and attack inputs without introducing heuristics and reliance on attack data.
However, training-free methods (e.g., layer-wise thresholds) often introduce new hand-crafted heuristics, and it is nearly impossible to manually anticipate the full spectrum of unforeseen scenarios~\cite{de2025comparative}, while supervised approaches typically rely on negative samples (i.e., attack data). To address these limitations, we formulate the task as unsupervised anomaly detection and focus on safe inputs.

\looseness=-1 \noindent \textbf{Formulation as anomaly detection.}
We train exclusively on safe inputs by framing detection as anomaly detection: the module learns the typical distribution of safety representations of safe inputs during training, and deviations at test time indicate potential attacks. Concretely, we introduce a Safety Pattern Auto‑Encoder (SPAE) trained to reconstruct safety representations of only safe inputs. At test time, inputs with low reconstruction error are classified as safe, while high error indicates attacks, consistent with anomaly detection literature~\cite{2014Anomaly}. We now detail SPAE's architecture, training, and testing.

\noindent \textbf{SPAE architecture.} SPAE uses a standard auto-encoder with a three-layer fully connected encoder and a symmetric decoder with ReLU activations. This hierarchical non-linear structure enables the model to capture complex inter-layer safety relations, considering that each dimension in safety representations corresponds to one layer. Formally, the encoder $f_{\text{enc}}$ compresses the safety representation $\mathbf{S}\text{r}$ into a low-dimensional latent space, and the decoder $f_{\text{dec}}$ reconstructs it back to the original dimensionality $|\mathcal{L}\text{s}|$:
\begin{equation}
\hat{\mathbf{S}_{\text{r}}} = f_{\text{dec}}\big(f_{\text{enc}}(\mathbf{S}_\text{r})\big).
\end{equation}

\noindent \textbf{SPAE training.}
The model is trained solely on safe inputs to minimize assumptions about attacks. The training objective minimizes the mean-squared reconstruction error between the input and reconstructed safety representations:
\begin{equation} \label{eq:loss}
\mathcal{L}_{\text{rec}} = \frac{1}{N_s} \sum_{i=1}^{N_s} || \hat{\mathbf{S}}_\text{r} - \mathbf{S}_\text{r} ||_2^2.
\end{equation}
Here, $N_s$ is the number of safe inputs. This loss encourages the SPAE to learn the latent distribution of safety representations. Therefore, it could detect attacks without reliance on attack data, ensuring generalization across diverse attacks.

\noindent \textbf{SPAE testing.} During testing, SPAE detects anomalies by evaluating the reconstruction error:
\begin{equation} \label{eq:error}
\delta = || \hat{\mathbf{S}}\text{r} - \mathbf{S}\text{r} ||_2^2.
\end{equation}
Given an input, if $\delta$ exceeds a threshold $\tau$, the input is flagged as attacked. Otherwise, the input is considered safe. Threshold selection is detailed in Appendix~\ref{appendix:threshold}.

\section{Experiments}
\label{sec:experiment}

\subsection{Experimental Setups}
\label{sec:exp_setup}

\noindent\textbf{Jailbreak attack methods.} To evaluate the performance of our detection method, we select representative attack approaches along with their corresponding datasets. For \emph{prompt-manipulation-based} attacks, we use: FC-Attack~\cite{zhang2025fc}, JOOD~\cite{jeong2025playing}, HADES~\cite{li2024images}, and MML~\cite{wang2025jailbreak}. For \emph{adversarial-perturbation-based} attacks, we employ VAJM~\cite{qi2024visual} and UMK~\cite{wang2024white}. Details of these attack methods are introduced in Appendix~\ref{appendix:jailbreak}.



 \noindent \textbf{Training and validation datasets.} 
To train LoD and select hyperparameters, we use unsafe inputs \(\mathbf{I}^-\) from AdvBench~\cite{zou2023universal}, which contains prompts describing unsafe behaviors. Safe inputs \(\mathbf{I}^+\) are from GQA~\cite{2019GQA}, a large-scale visual reasoning dataset for compositional question answering. We combine prompts from both datasets and create multimodal inputs by synthesizing images for both unsafe and safe texts using Pixart-Sigma~\cite{chen2024pixart}.
For the MSCAV classifiers, we allocate 100 safe-unsafe pairs for training and 100 for validation. The autoencoder is trained exclusively on 320 safe inputs, with 80 reserved for validation.

\looseness=-1 \noindent \textbf{Testing datasets.}
The \emph{test set} consists of 400 safe inputs from MM-Vet2~\cite{yu2024mm} and 400 unsafe inputs from MM-SafetyBench~\cite{liu2024mm}. Approximately 100 remaining safe inputs are held out for threshold selection (see Appendix~\ref{appendix:threshold}). For attack methods other than HADES, attacked samples are generated by applying each attack method to the unsafe inputs from MM-SafetyBench, which cover 13 safety-related scenarios. HADES is treated separately because it provides its own dataset, covering five categories of prohibited questions.
We further provide additional experiments with an expanded dataset size in Appendix~\ref{appendix:expanded}

\looseness=-1 \noindent\textbf{Baselines.} To ensure a comprehensive evaluation and highlight the strengths of our proposed method, we select representative baselines from two main categories of detection approaches:
(a) For methods that do not use internal representations, we include CIDER~\cite{xu2024cross}, MirrorCheck~\cite{fares2024mirrorcheck}, and JailGuard~\cite{zhang2025jailguard};
(b) For methods leveraging internal representations, we incorporate HiddenDetect~\cite{jiang2025hiddendetect} and GradSafe~\cite{xie2024gradsafe}. 
For GradSafe, we use GradSafe-Zero instead of GradSafe-Adapt, as the latter requires training on attack data, contrary to our experiment settings.

\noindent\textbf{Evaluation criteria.}  
We follow the prior work~\cite{jiang2025hiddendetect} and evaluate detection accuracy using AUROC, which reflects performance across all decision thresholds. Results on additional criteria, including F1 and TPR at low FPR (e.g., FPR 
$\leq$ 0.01), are reported in Appendix~\ref{Appendix:eva}. We also report the computation time in seconds to analyze detection efficiency.

\looseness=-1 \noindent\textbf{LVLMs.}  
We evaluate our method on three LVLMs: LLaVA-1.6-vicuna-7B~\cite{liu2024llavanext}, CogVLM-chat-v1.1~\cite{wang2023cogvlm}, and Qwen2.5-VL-7B-Instruct~\cite{Qwen2.5-VL}.

\noindent \textbf{Hyperparameter settings.} 
Across all experiments, we set $P_0 = 0.9$. While AUROC is threshold-independent, specific LVLM thresholds are detailed in Appendix~\ref{appendix:threshold}, with F1 scores reported across various settings. Baseline configurations are provided in Appendix~\ref{appendix:baselines}.

\subsection{Overall Detection Performance}
\label{sec:results} 
Table~\ref{tab:comparison} reports a rigorous performance comparison across various LVLMs and six jailbreak attacks using the AUROC metric. From the results, several key observations emerge:

First, our proposed LoD framework consistently achieves the highest overall performance, outperforming the strongest baselines in average AUROC by 19.32\%, 4.50\%, and 17.11\% across the three LVLMs, respectively. Most notably, LoD substantially elevates the worst-case performance (i.e., the minimum AUROC), with remarkable relative gains of 170.4\%, 3.94\%, and 26.57\%. Although LoD yields to GradSafe on the MML dataset, it consistently maintains the second-best performance with a marginal gap, and we try to explain this with a case study in Appendix~\ref {appendix:boundary_case}.

Second, we observe a significant performance gap between internal and non-internal methods. Baselines leveraging internal representations generally yield higher average AUROC than methods based on inputs or outputs. However, their Min column reveals a critical weakness: hand-crafted heuristics can be overly specialized. When an attack pattern deviates from their predefined rules, these methods often fail. For instance, while GradSafe is highly effective on MML, its performance drops to near-random levels on UMK. In contrast, LoD achieves superior detection accuracy across diverse adversarial scenarios.
Furthermore, supplementary analysis in Appendix~\ref{appendix:ood_benign} suggests that LoD remains robust to complex, benign out-of-distribution (OOD) prompts by consistently maintaining low anomaly scores.

\begin{figure}[t]
\centering
\includegraphics[width=1\columnwidth]{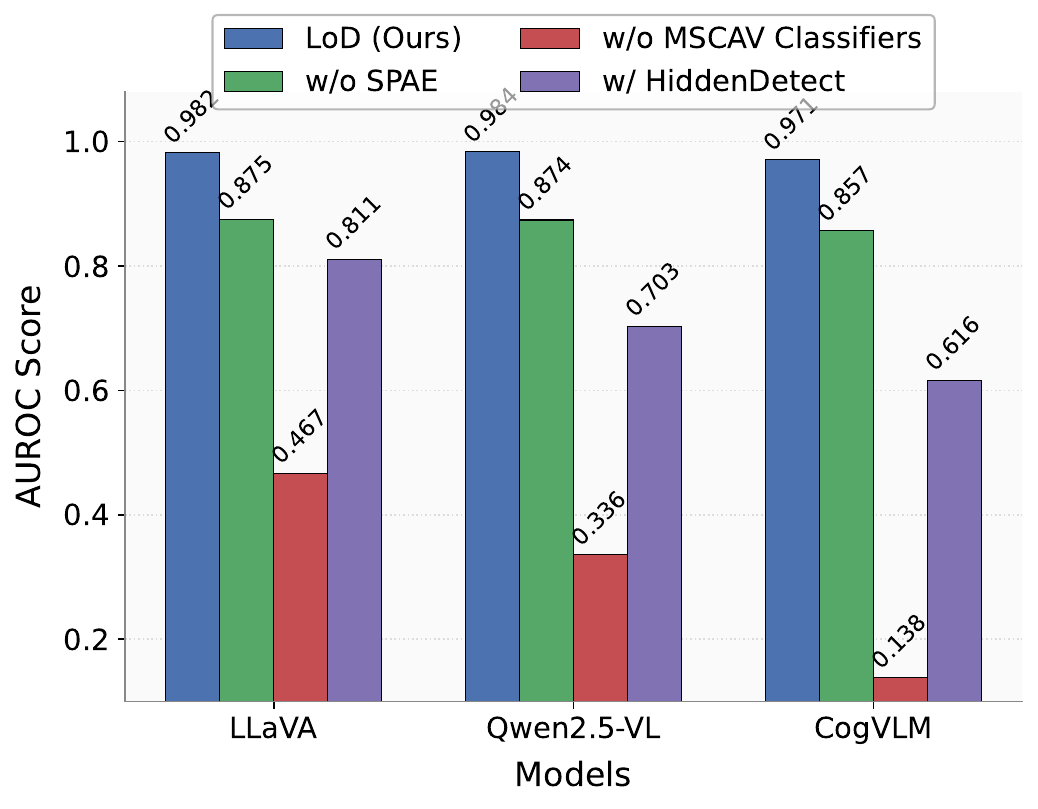}
\caption{
\looseness=-1 Ablation study results showing the average AUROC scores on three LVLMs across six attack methods under different ablation settings. 
}
\label{fig:ablation}
\end{figure}

\subsection{Ablation Study}
\label{subsec:ablation}
To better understand the contribution of each core component in the LoD framework, we compare our method with the following ablated variants:
\begin{itemize}[nosep,leftmargin=1em,labelwidth=*,align=left]
 \item \textbf{LoD w/o MSCAV Classifiers}: This variant evaluates the contribution of the representation learning module by replacing safety representations with high-dimensional raw internal activations concatenated across all layers.

    \item  \textbf{LoD w/o SPAE}: This variant demonstrates the effectiveness of SPAE by replacing it with a simple aggregation of safety representations: an input is flagged as unsafe if any dimension of the refined safety representation \(\mathbf{S}_\text{r}\) exceeds its corresponding threshold. Specifically, the threshold for each dimension is defined as the 95th percentile of the safety representations observed in the held-out validation set from MM-Vet2.

    \item \textbf{LoD w/ HiddenDetect}:  This variant compares our MSCAV-based safety representation learning with that in an existing state-of-the-art method, HiddenDetect. Specifically, we replace our safety representations with the feature extracted from the method in HiddenDetect.
\end{itemize}       
Additional results in Appendix~\ref{appendix:spae_unsafe} show  the effect of introducing unsafe data into SPAE training, which brings no consistent gains.

\looseness=-1 Results in Figure~\ref{fig:ablation} confirm that both MSCAV classifiers and SPAE are indispensable. Removing the MSCAV classifiers (\textit{w/o MSCAV Classifiers}) triggers the most severe performance collapse, validating its importance in extracting discriminative safety features from the raw activations. While incorporating HiddenDetect features (w/ HiddenDetect) partially restores performance, it still trails the full LoD model by a significant margin (e.g., AUROC from 0.811 vs. 0.982), suggesting safety representation's superior generalization across LVLMs. Similarly, the consistent drop in the w/o SPAE variant (down to 0.857--0.875 AUROC) demonstrates that simply aggregating safety representation dimensions is insufficient and that the SPAE module is necessary. 

\begin{table}[htbp]
\centering
\footnotesize 
\setlength{\tabcolsep}{8pt} 
\renewcommand{\arraystretch}{1.2} 
\begin{tabular}{lccc}
\toprule
\textbf{Attack} & \textbf{LLaVA} & \textbf{Qwen2.5-VL} & \textbf{CogVLM} \\
\midrule
FC-Attack       & 45.34\%        & 28.92\%          & 52.44\%         \\
JOOD            & 61.53\%        & 52.07\%          & 61.91\%         \\
HADES           & 23.03\%        & 35.75\%          & 21.65\%         \\
MML             & 49.85\%        & 52.38\%          & 52.13\%         \\
VAJM            & 32.26\%        & 31.35\%          & 36.78\%         \\
UMK             & 52.83\%        & 44.08\%          & 34.98\%         \\
\midrule
\textbf{Average} & 44.14\% & 40.76\%& 43.32\%\\
\bottomrule
\end{tabular}
\caption{Average overlap ratios between safe and attacked safety representations across dimensions.}
\label{tab:overlap}
\end{table}

To understand why removing SPAE degrades performance, we analyze the per-dimension characteristics of the learned safety representations. Despite clear separability in average trends (Figure~\ref{fig:pattern1}), a fine-grained, per-dimension analysis reveals substantial overlap. We quantify this by defining an overlapping range per dimension between the extreme values of attacked (min) and safe (max) samples. As shown in Table~\ref{tab:overlap}, the average overlap ratio consistently exceeds 40\% across models. This high degree of overlap renders direct per-dimension thresholds ineffective for reliable separation, thus necessitating the SPAE module to 
project these vectors into a discriminative one-dimensional space.

\subsection{Parameter Sensitivity Analysis}

Our method involves two hyperparameters: the layer selection threshold $P_0$ and the decision threshold $\tau$. 
Because AUROC is computed by integrating over all possible $\tau$ values and represents an average performance metric, it is invariant to the choice of $\tau$.  
We therefore focus on analyzing the impact of $P_0$ here.
We further discuss the effect of $\tau$ on threshold-dependent criteria (e.g., F1-score) in Appendix~\ref{appendix:threshold}, and analyse the sensitivity to the unsafe training dataset in Appendix~\ref{appendix:harmbench_sensitivity}.

\begin{table}[htbp]
\centering
\footnotesize
\setlength{\tabcolsep}{4pt}
\renewcommand{\arraystretch}{1.2}
\begin{NiceTabular*}{0.9\columnwidth}{@{\extracolsep{\fill}} l c c c @{}} 
\toprule
\multirow{2}{*}{\textbf{$P_0$}} & \multicolumn{3}{c}{\textbf{Average AUROC over six attacks}} \\
\cmidrule(lr){2-4}
& \textbf{LLaVA} & \textbf{Qwen2.5-VL} & \textbf{CogVLM} \\
\midrule
0.80 & 0.9821 & 0.9837 & 0.9704 \\
\rowcolor{gray!20}
0.90 & 0.9820 & 0.9838 & 0.9705 \\
0.95 & 0.9815 & 0.9837 & 0.9840 \\
0.97 & 0.9821 & 0.9837 & 0.9840 \\
0.99 & 0.9743 & 0.9896 & 0.9837 \\
\bottomrule
\end{NiceTabular*}
\caption{Average AUROC scores over six attack methods for different $P_0$ thresholds across three LVLMs. 
The threshold we used is highlighted in gray. 
}
\label{tab:sensitivity}
\end{table}

\looseness=-1 As shown in  Table~\ref{tab:sensitivity},  the average AUROC remains consistently high for all LVLMs and $P_0$ values, indicating that LoD is highly robust to the layer selection threshold. Moderate $P_0$ values (e.g., 0.90) yield uniformly strong results, while overly large $P_0$ (e.g., 0.99) may cause minor degradation on LLaVA due to the removal of informative layers. 

\begin{table}[htbp]
\centering
\footnotesize
\setlength{\tabcolsep}{4pt}
\begin{tabular}{lccc}
\toprule
\textbf{Method} & \textbf{LLaVA} & \textbf{Qwen2.5-VL} & \textbf{CogVLM} \\
\midrule
MirrorCheck     & 1.99 & 2.16 & 3.08 \\
CIDER           & 0.91 & \underline{0.75} & 1.38 \\
JailGuard       & 15.22 & 21.24 & 16.39 \\
GradSafe        & 0.85 & 0.87$^\dagger$ & 1.08 \\
HiddenDetect    & \underline{0.67} & 0.86 & \underline{0.39} \\
\midrule
\rowcolor{gray!10}
\textbf{Ours}    & \textbf{0.25} & \textbf{0.62} & \textbf{0.34} \\
\midrule
$\Delta$    & $-62.7\%$ & $-17.3\%$ & $-12.8\%$ \\
\bottomrule
\end{tabular}

\caption{
\looseness=-1
Average detection time (in seconds) per input across three LVLMs. 
Best and second  best results are shown in \textbf{bold} and \underline{underlined}, respectively.
$\Delta$ denotes the relative change in detection time of LoD over the best baseline.
}
\label{tab:detection_time}
\end{table}

\subsection{Computational Efficiency Analysis}

As shown in Table~\ref{tab:detection_time}, our LoD method is the most efficient across three LVLMs, with a computation time ranging from 0.25 to 0.62 seconds. Here, most methods are run on a single NVIDIA A800 GPU, except for GradSafe on Qwen2.5-VL, which is run on 4 GPUs due to high memory usage.
As shown in the table, LoD reduces detection time by up to {62.7\%} compared to the best baseline on LLaVA, and shows improvements of {17.3\%} and {12.8\%} on Qwen2.5-VL and CogVLM, respectively. 
In terms of training cost, fitting our MSCAV classifiers and the SPAE module requires only 12 and 2 seconds, respectively, which is a one-time offline training. These results highlight the efficiency of LoD. A detailed breakdown of these costs is provided in Appendix~\ref{appendix:overhead}.


\section{Conclusion}
In this work, we introduce Learning to Detect, a learnable framework that leverages LVLMs' internal activations for jailbreak detection. While existing methods struggle to achieve both high accuracy and generalization simultaneously, LoD solves this by training task-specific parameters on safe and unsafe data, eliminating reliance on both attack-specific data and hand-crafted heuristics.  Experiments show that LoD consistently outperforms state-of-the-art baselines across multiple LVLMs and previously unseen attacks.

\section*{Limitations}
\looseness=-1  Although our method achieves consistent performance across models and attacks, it has several limitations. 
First, 
its results depend on the specific LVLM used.
Major architectural changes or new safety alignment mechanisms may require adapting the detector, though the prevalence of deep neural architectures suggests our approach will remain broadly applicable. Second, while our method is demonstrated to be effective against diverse jailbreaks, our evaluation cannot fully anticipate future adaptive attacks that specifically manipulate internal representations. 
Exploring such scenarios is an important step toward building more resilient defenses.



\bibliography{custom}

\appendix

\section{Supplementary Evaluation Using Additional Criteria}
\label{Appendix:eva}
\looseness=-1 To further validate the effectiveness and robustness of our detection method, we note that the main text already demonstrates high AUROC scores. However, AUROC is an average criterion that may not fully reflect performance under specific operating conditions. In this section, we pursue two objectives: first, we measure TPR at low FPR to evaluate detection capability under strict false positive constraints; second, we determine practical thresholds for our method and verify its effectiveness at these thresholds.

\subsection{TPR at Low FPR}  
\label{appendix:tpr}

\begin{table*}[t]
\centering
\fontsize{8.5pt}{10.2pt}\selectfont 
\begin{tabular*}{\textwidth}{@{\extracolsep{\fill}} l l c c c c c c c  @{}}
\toprule
\shortstack[l]{\vspace{-0.8em}\textbf{Model}} & 
\shortstack[l]{\vspace{-0.8em}\textbf{Method}} & 
\multicolumn{6}{c}{\shortstack[l]{\vspace{0em}\textbf{Results on Jailbreak Attacks (TPR @ FPR=0.01)}}} & 
\shortstack[l]{\vspace{-1.0em}\textbf{Avg.}} \\
\cmidrule(lr){3-8}
 & & \multicolumn{1}{c}{FC-Attack} & \multicolumn{1}{c}{JOOD} & \multicolumn{1}{c}{HADES} & \multicolumn{1}{c}{MML} & \multicolumn{1}{c}{VAJM} & \multicolumn{1}{c}{UMK}  \\
\midrule
\multirow{7}{*}{LLaVA}
    & MirrorCheck$^{(a)}$      & 0.007 & 0.003 & 0.000 & 0.000 & 0.015 & 0.000 & 0.004 \\
    & CIDER$^{(a)}$            & \underline{0.273} & 0.010 & 0.000 & 0.000 & 0.000 & 0.000 & 0.047 \\
    & JailGuard$^{(a)}$        & 0.000 & 0.007 & 0.003 & 0.005 & 0.007 & 0.010 & 0.005 \\
    & GradSafe$^{(b)}$         & 0.000 & 0.045 & 0.015 & \textbf{0.477} & 0.000 & 0.000 & 0.090 \\
    & HiddenDetect$^{(b)}$     & 0.000 & \underline{0.295} & \underline{0.865} & 0.000 & \underline{0.430} & \underline{0.003} & \underline{0.265} \\
    & Ours$^{(b)}$             & \textbf{0.477} & \textbf{0.443} & \textbf{0.980} & 0.000 & \textbf{0.797} & \textbf{0.223} & \textbf{0.487} \\
    \cmidrule(lr){2-9}
    & $\Delta$                 & +74.7\% & +50.2\% & +13.3\% & -100\% & +85.3\% & +7333\% & +83.8\% \\
\midrule
\multirow{6}{*}{Qwen2.5}
    & MirrorCheck$^{(a)}$      & 0.003 & 0.000 & 0.000 & 0.000 & 0.000 & 0.000 & 0.000 \\
    & CIDER$^{(a)}$            & 0.412 & 0.048 & 0.007 & 0.000 & 0.000 & 0.000 & 0.078 \\
    & JailGuard$^{(a)}$        & 0.000 & 0.000 & 0.000 & 0.000 & 0.003 & 0.003 & 0.001 \\
    & GradSafe$^{(b)}$         & \underline{0.802} & \underline{0.618} & \underline{0.718} & \textbf{0.095} & \underline{0.890} & \underline{0.757} & \underline{0.647} \\
    & HiddenDetect$^{(b)}$     & 0.000 & 0.172 & 0.410 & 0.000 & 0.230 & 0.000 & 0.135 \\
    & Ours$^{(b)}$             & \textbf{0.999} & \textbf{0.815} & \textbf{0.983} & 0.000 & \textbf{0.995} & \textbf{0.790} & \textbf{0.764} \\
    \cmidrule(lr){2-9}
    & $\Delta$                 & +24.6\% & +31.9\% & +36.9\% & -100\% & +11.8\% & +4.4\% & +18.1\% \\
\midrule
\multirow{6}{*}{CogVLM}
    & MirrorCheck$^{(a)}$      & 0.000 & 0.007 & 0.000 & 0.003 & 0.007 & 0.000 & 0.003 \\
    & CIDER$^{(a)}$            & \underline{0.122} & 0.022 & 0.000 & 0.000 & 0.000 & 0.000 & 0.024 \\
    & JailGuard$^{(a)}$        & 0.000 & 0.010 & 0.018 & 0.000 & 0.058 & 0.030 & 0.019 \\
    & GradSafe$^{(b)}$         & 0.000 & 0.003 & 0.000 & \textbf{1.000} & 0.050 & 0.003 & \underline{0.176} \\
    & HiddenDetect$^{(b)}$     & 0.000 & \underline{0.080} & \underline{0.328} & 0.000 & \underline{0.163} & \underline{0.175} & 0.124 \\
    & Ours$^{(b)}$             & \textbf{0.705} & \textbf{0.767} & \textbf{0.998} & 0.000 & \textbf{0.938} & \textbf{0.895} & \textbf{0.717} \\
    \cmidrule(lr){2-9}
    & $\Delta$                 & +477.9\% & +858.8\% & +204.3\% & -100\% & +475.5\% & +411.4\% & +307.4\% \\
\bottomrule
\end{tabular*}
\caption{TPR at FPR=0.01 across different LVLMs and detection methods. Methods are categorized into two categories based on whether they utilize only the input and output information of LVLMs$^{(a)}$ or utilize information of internal representations in the LVLMs$^{(b)}$. The safe-test data are sampled from the held-out set of MM-Vet2. The best results are shown in \textbf{bold}, and the second-best results are shown in \underline{underline}. $\Delta$ denotes the relative improvement of our method over the best baseline.}
\label{tab:tpr_lowfpr}
\end{table*}



Table~\ref{tab:tpr_lowfpr} reports the TPR at a strictly fixed FPR of 0.01. Most baseline methods, particularly those utilizing only input-output information$^{(a)}$, exhibit very low TPR, often failing to detect any attacks. This stems from the stringent FPR threshold, and most baselines cannot provide sufficient discriminative power to separate jailbreaks from safe inputs at this granularity. 

While representation-based baselines like GradSafe and HiddenDetect show improved performance, their effectiveness is highly volatile. For instance, GradSafe excels on the MML dataset but struggles significantly with HADES and UMK, whereas HiddenDetect shows the opposite trend. In contrast, our method, LoD, consistently achieves the highest average TPR across all three models, with a particularly remarkable improvement on CogVLM, where it outperforms the best baseline by over 300\%. Although LoD encounters challenges with the MML attack, which appears to follow a distinct distribution better captured by gradient-based methods like GradSafe, it maintains a dominant lead in detecting attacks such as HADES, VAJM, and UMK. These results underscore the robustness and superior generalization of LoD under practical, low-FPR operating conditions.

\subsection{Threshold Decision}
\label{appendix:threshold}
\begin{figure*}[t]
\centering
\includegraphics[width=2\columnwidth]{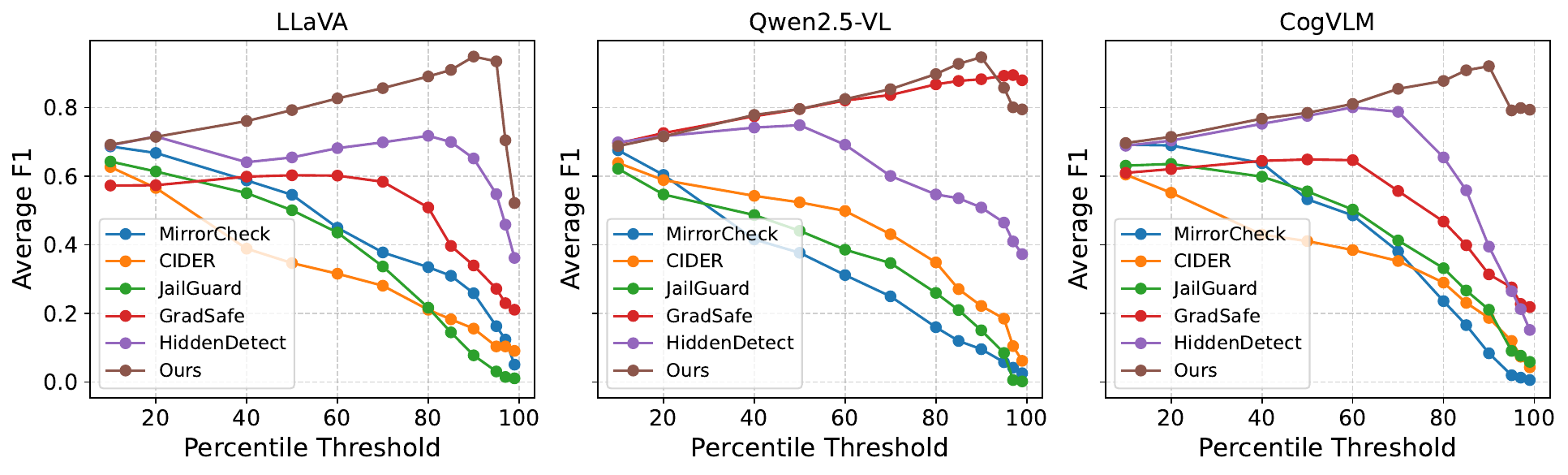}
\caption{F1 score sensitivity of different detection methods on the three models across threshold percentiles from 10th to 99th. Our method consistently achieves the highest F1 at each percentile, demonstrating robustness to threshold selection.}

\label{fig:f1-sensitivity}
\end{figure*}

\begin{table*}[t]
\centering
\fontsize{9pt}{10.8pt}\selectfont 

\begin{tabular*}{\textwidth}{@{\extracolsep{\fill}} l l c c c c c c c @{}}
\toprule
\shortstack[l]{\vspace{-0.8em}\textbf{Model}} & 
\shortstack[l]{\vspace{-0.8em}\textbf{Method}} & 
\multicolumn{6}{c}{\shortstack[l]{\vspace{-0em}\textbf{Results on Jailbreak Attacks (F1 Score)}}} & 
\shortstack[l]{\vspace{-1em}\textbf{Average}} \\
\cmidrule(lr){3-8}
 &  & FC-Attack & JOOD & HADES & MML & VAJM & UMK &  \\
\midrule
\multirow{7}{*}{LLaVA}
    & MirrorCheck$^{(a)}$      & 0.335 & 0.174 & 0.034 & 0.067 & 0.869 & 0.075 & 0.259 \\
    & CIDER$^{(a)}$            & 0.544 & 0.285 & 0.100 & 0.000 & 0.005 & 0.000 & 0.156 \\
    & JailGuard$^{(a)}$        & 0.009 & 0.100 & 0.152 & 0.023 & 0.120 & 0.066 & 0.078 \\
    & GradSafe$^{(b)}$         & 0.234 & 0.355 & 0.430 & \underline{0.947} & 0.074 & 0.000 & 0.340 \\
    & HiddenDetect$^{(b)}$     & \underline{0.741} & \underline{0.802} & \underline{0.943} & 0.000 & \underline{0.926} & \underline{0.497} & \underline{0.652} \\
    & Ours$^{(b)}$             & \textbf{0.960} & \textbf{0.908} & \textbf{0.963} & \textbf{0.955} & \textbf{0.957} & \textbf{0.953} & \textbf{0.949} \\
    \cmidrule(lr){2-9}
    & $\Delta$                 & +29.6\% & +13.2\% & +2.1\% & +0.8\% & +3.3\% & +91.8\% & +45.6\% \\
\midrule
\multirow{6}{*}{Qwen2.5}
    & MirrorCheck$^{(a)}$      & 0.325 & 0.070 & 0.045 & 0.132 & 0.005 & 0.000 & 0.096 \\
    & CIDER$^{(a)}$            & 0.826 & 0.267 & 0.081 & 0.000 & 0.151 & 0.009 & 0.222 \\
    & JailGuard$^{(a)}$        & 0.289 & 0.073 & 0.225 & 0.000 & 0.185 & 0.133 & 0.151 \\
    & GradSafe$^{(b)}$         & \underline{0.892} & 0.813 & 0.847 & \underline{0.947} & \underline{0.921} & \underline{0.876} & \underline{0.883} \\
    & HiddenDetect$^{(b)}$     & 0.033 & \underline{0.876} & \underline{0.896} & 0.000 & 0.734 & 0.517 & 0.509 \\
    & Ours$^{(b)}$             & \textbf{0.955} & \textbf{0.939} & \textbf{0.957} & \textbf{0.951} & \textbf{0.951} & \textbf{0.931} & \textbf{0.947} \\
    \cmidrule(lr){2-9}
    & $\Delta$                 & +7.1\% & +7.2\% & +6.8\% & +0.4\% & +3.3\% & +6.3\% & +7.2\% \\
\midrule
\multirow{6}{*}{CogVLM}
    & MirrorCheck$^{(a)}$      & 0.108 & 0.104 & 0.005 & 0.108 & 0.128 & 0.049 & 0.084 \\
    & CIDER$^{(a)}$            & \underline{0.649} & \underline{0.388} & 0.057 & 0.000 & 0.027 & 0.000 & 0.187 \\
    & JailGuard$^{(a)}$        & 0.000 & 0.201 & 0.216 & 0.000 & 0.508 & 0.344 & 0.211 \\
    & GradSafe$^{(b)}$         & 0.000 & 0.114 & 0.065 & \textbf{0.947} & 0.558 & 0.202 & 0.314 \\
    & HiddenDetect$^{(b)}$     & 0.009 & 0.373 & \underline{0.748} & 0.000 & \underline{0.633} & \underline{0.606} & \underline{0.395} \\
    & Ours$^{(b)}$             & \textbf{0.931} & \textbf{0.886} & \textbf{0.941} & \underline{0.906} & \textbf{0.928} & \textbf{0.933} & \textbf{0.921} \\
    \cmidrule(lr){2-9}
    & $\Delta$                 & +43.5\% & +128.4\% & +25.8\% & -4.3\% & +46.6\% & +54.0\% & +133.2\% \\
\bottomrule
\end{tabular*}
\caption{F1 scores across different models and methods. Methods are categorized into two categories based on whether they utilize only the input and output information of LVLMs$^{(a)}$ or  utilize information of internal representations in the LVLMs$^{(b)}$. The safe-test data are sampled from the held-out set of MM-Vet2, and the detection threshold is determined based on the 90th percentile of the anomaly score distribution. Best results among baselines are shown in \textbf{bold}, second-best baseline results are shown in \underline{underline}. $\Delta$ denotes the relative improvement of our method over the best baseline.}
\label{tab:f1}
\end{table*}


To determine thresholds for applying our method in real-world scenarios, we use the held-out set from the MM-Vet2~\cite{yu2024mm} dataset, which contains approximately 100 items, each consisting of a question for VLMs paired with a corresponding image.
 It serves as a collection of safe samples, which we use to align the measurement scores and select the 90th percentile as the threshold, thereby reducing the false positive rate. The resulting thresholds are 0.25308 for LLaVA, 0.33530 for Qwen2.5-VL, and 0.20748 for CogVLM. We apply the same procedure to determine thresholds for all baselines.

\looseness=-1 To evaluate the effectiveness of the chosen threshold, we report F1 scores across models and datasets in Table~\ref{tab:f1}. Our method consistently achieves high values, significantly outperforming all baselines. For LLaVA, we improve over the best baseline by +45.6\% on average, with especially large gains on UMK (+91.8\%). Similar improvements hold for Qwen2.5-VL (+7.2\% on average) and CogVLM (+133.2\%), indicating that the gains are not tied to a single model but are robust across architectures.  
Notably, while baselines such as GradSafe or HiddenDetect achieve strong performance on certain datasets but collapse on others (often dropping to near zero), our method maintains consistently high F1 across all settings.
Such inconsistencies suggest that thresholds determined by these baselines lack generalizability: their score measurements for different safe samples vary considerably, leading to unstable and limited detection performance.
 By contrast, our method maintains consistently high F1 across all settings, without significant degradation on any benchmark. This demonstrates not only that the threshold selected generalizes well across different models and attacks, but also that it remains effective under complex attack scenarios.

\noindent \textbf{Sensitivity analysis.} To evaluate the impact of percentile selection on practical detection performance, we investigate the variation of average F1 scores across a range from the 10th to the 99th percentile. Figure~\ref{fig:f1-sensitivity} presents the comparative results. While the F1 scores of all methods fluctuate to some extent as the percentile varies, our method consistently achieves the highest detection performance, with the sole exception being on Qwen2.5-VL, where it is slightly surpassed by GradSafe from the 95th percentile onwards. Nonetheless, in the vast majority of cases, LoD maintains state-of-the-art (SOTA) performance. This demonstrates the superior robustness of our approach and its capability to provide stable detection efficacy across a wide range of operational configurations.

\section{Expanded-Set Robustness Check}
\label{appendix:expanded}

To further evaluate whether the main conclusions remain stable under a larger evaluation scale, we conduct an expanded-set robustness check. The controlled benchmark in Table~\ref{tab:comparison} is constructed by random sampling from datasets that already cover multiple prohibited scenarios. Therefore, this experiment is intended primarily to test robustness to evaluation scale and sampling variation, rather than to introduce a new scenario-coverage benchmark or replace the controlled comparison.

\noindent\textbf{Expanded data.}
The safe set uses all 517 samples from MM-Vet2. For attack inputs, we use the largest available evaluation pool for each attack. Specifically, we use the full released datasets for  HADES, containing 750 samples. For FC-Attack, JOOD, MML, VAJM, and UMK, we generate additional attack inputs from MM-SafetyBench~\cite{liu2024mm}. This yields more than 2,300 MML samples and more than 1,600 samples for JOOD, VAJM, and UMK. Compared with the controlled benchmark, the expanded evaluation mainly increases the number of test instances within the existing scenario taxonomy, while keeping the training data, validation data, and detector hyperparameters unchanged. Since AUROC is computed from pairwise rankings between benign and attacked inputs and is insensitive to the decision threshold, the different numbers of benign and attacked samples do not affect the validity of comparing detection performance across methods.

\noindent\textbf{Compared methods.}
Due to the substantially increased evaluation scale, running generation-heavy baselines on the full expanded sets would incur substantial computational cost. For example, JailGuard~\cite{zhang2025jailguard} requires multiple LVLM generations for each input, which becomes impractical at this scale. Therefore, this large-scale robustness check focuses on GradSafe and HiddenDetect, the two strongest internal-representation baselines in the controlled benchmark. The complete comparison with all baselines is reported in Table~\ref{tab:comparison}.

\begin{table*}[htbp]
\centering
\fontsize{9pt}{10.8pt}\selectfont 

\begin{tabular*}{\textwidth}{@{\extracolsep{\fill}} l l c c c c c c c @{}}
\toprule
\shortstack[l]{\vspace{-0.8em}\textbf{Model}} & 
\shortstack[l]{\vspace{-0.8em}\textbf{Method}} & 
\multicolumn{6}{c}{\shortstack[l]{\vspace{-0em}\textbf{Results on Expanded Attack Sets (AUROC)}}} & 
\shortstack[l]{\vspace{-1em}\textbf{Average}} \\
\cmidrule(lr){3-8}
 & & FC-Attack & JOOD & HADES & MML & VAJM & UMK & \\
\midrule
\multirow{4}{*}{LLaVA}
    & GradSafe      & 0.7869 & 0.6485 & 0.8282 & \textbf{0.9969} & 0.5564 & 0.0318 & 0.6415 \\
    & HiddenDetect  & \underline{0.9246} & \underline{0.9002} & \underline{0.9919} & 0.5785 & \underline{0.9669} & \underline{0.8433} & \underline{0.8676} \\
    & Ours          & \textbf{0.9927} & \textbf{0.9654} & \textbf{0.9992} & \underline{0.9746} & \textbf{0.9956} & \textbf{0.9699} & \textbf{0.9829} \\
    \cmidrule(lr){2-9}
    & $\Delta$      & +7.37\% & +7.24\% & +0.74\% & -2.24\% & +2.97\% & +15.01\% & +13.29\% \\
\midrule
\multirow{4}{*}{Qwen2.5-VL}
    & GradSafe      & \underline{0.6043} & 0.8700 & 0.8944 & \textbf{0.9918} & \underline{0.9625} & \underline{0.9316} & \underline{0.8758} \\
    & HiddenDetect  & 0.4350 & \underline{0.9201} & \underline{0.9638} & 0.6160 & 0.8888 & 0.8569 & 0.7801 \\
    & Ours          & \textbf{1.0000} & \textbf{0.9867} & \textbf{0.9994} & \underline{0.9293} & \textbf{1.0000} & \textbf{0.9895} & \textbf{0.9841} \\
    \cmidrule(lr){2-9}
    & $\Delta$      & +65.48\% & +7.24\% & +3.69\% & -6.30\% & +3.90\% & +6.22\% & +12.37\% \\
\midrule
\multirow{4}{*}{CogVLM}
    & GradSafe      & 0.6780 & 0.3417 & 0.4024 & \textbf{1.0000} & 0.7604 & 0.6522 & 0.6391 \\
    & HiddenDetect  & \underline{0.8855} & \underline{0.7607} & \underline{0.9065} & 0.7515 & \underline{0.8293} & \underline{0.8563} & \underline{0.8316} \\
    & Ours          & \textbf{0.9921} & \textbf{0.9673} & \textbf{0.9975} & \underline{0.9028} & \textbf{0.9893} & \textbf{0.9698} & \textbf{0.9698} \\
    \cmidrule(lr){2-9}
    & $\Delta$      & +12.04\% & +27.16\% & +10.04\% & -9.72\% & +19.29\% & +13.25\% & +16.62\% \\
\bottomrule
\end{tabular*}
\caption{AUROC results on expanded attack sets. The best results are shown in \textbf{bold}, and the second-best results are \underline{underlined}. $\Delta$ denotes the relative improvement of our method over the best baseline.}
\label{tab:expanded_auroc}
\end{table*}

As shown in Table~\ref{tab:expanded_auroc}, LoD remains effective under the expanded evaluation setting, achieving average AUROC scores of 0.9829, 0.9841, and 0.9698 on LLaVA, Qwen2.5-VL, and CogVLM, respectively. These results are consistent with the controlled benchmark, indicating that LoD's advantage is not an artifact of a small test set. Although GradSafe remains strong on MML, its performance is less stable across other attacks, especially on LLaVA and CogVLM. HiddenDetect is competitive on several attacks but still achieves a lower average AUROC than LoD across all three LVLMs. These results suggest that LoD remains robust when covering larger attack pools.

\section{Benign OOD Scoring Analysis}
\label{appendix:ood_benign}

To examine how SPAE scoring behaves on complex but entirely benign instructions that are topically distant from the safety-oriented training data, we evaluate benign OOD prompts from MMMU~\cite{yue2023mmmu}. We select three representative subject groups: art, computer science, and mathematics, covering creative, technical, and symbolic reasoning inputs. This analysis focuses on the distribution of SPAE anomaly scores rather than constructing a new attack-detection benchmark.

\begin{table*}[t]
\centering
\fontsize{8.5pt}{10.2pt}\selectfont
\begin{tabular*}{\textwidth}{@{\extracolsep{\fill}} l c c c c c c @{}}
\toprule
\textbf{Model} & 
\textbf{MM-Vet2} & 
\textbf{Art} & 
\textbf{CS} & 
\textbf{Math} & 
\textbf{Attack Range} \\
\midrule
LLaVA
& 0.1044$\pm$0.1343 
& 0.0290$\pm$0.0398 
& 0.1022$\pm$0.0834 
& 0.1132$\pm$0.0752 
& 0.4804--0.8971 $(\sigma: 0.0142$--$0.2384)$ \\
Qwen2.5-VL 
& 0.1046$\pm$0.1345 
& 0.0624$\pm$0.0688 
& 0.1268$\pm$0.0869 
& 0.1336$\pm$0.0831 
& 0.3824--0.8343 $(\sigma: 0.0099$--$0.1705)$ \\
CogVLM 
& 0.0787$\pm$0.0982 
& 0.0330$\pm$0.0407 
& 0.0783$\pm$0.0546 
& 0.1051$\pm$0.0619 
& 0.2248--0.5394 $(\sigma: 0.0113$--$0.1427)$ \\
\bottomrule
\end{tabular*}
\caption{SPAE anomaly score distributions on safe test data, benign OOD prompts, and attack inputs. Lower scores indicate closer alignment with the learned safe manifold. The attack range reports the range of mean anomaly scores across jailbreak attack sets, with the corresponding range of standard deviations shown in parentheses.}
\label{tab:ood_benign_scores}
\end{table*}

As shown in Table~\ref{tab:ood_benign_scores}, the benign OOD prompts yield anomaly scores comparable to the MM-Vet2 and substantially below the attack-score range across all three LVLMs. This indicates that SPAE does not simply assign high anomaly scores to complex benign prompts that are topically distant from the safety training data.

\section{Effect of Unsafe Representations in SPAE Training}
\label{appendix:spae_unsafe}
To isolate the effect of incorporating unsafe representations into SPAE training, we train an additional SPAE variant alongside the original one-class SPAE used in LoD. The original SPAE is trained only on safe multimodal representations, whereas SPAE-unsafe is trained with the same safe representations plus vanilla unsafe representations from AdvBench using an additional margin-based penalty on unsafe reconstruction.
Specifically, SPAE-unsafe optimizes the following objective:
\[
\mathcal{L}_{\mathrm{unsafe}}
=
\mathrm{MSE}_{\mathrm{safe}}
+
\lambda \max\left(0, M - \mathrm{MSE}_{\mathrm{unsafe}}\right),
\]
where \(\mathrm{MSE}_{\mathrm{safe}}\) denotes the reconstruction error on safe representations, \(\mathrm{MSE}_{\mathrm{unsafe}}\) denotes the reconstruction error on unsafe representations, \(M=2.0\) is the margin, and \(\lambda=0.1\) controls the penalty strength.

\begin{table*}[htbp]
\centering
\fontsize{8.5pt}{10.2pt}\selectfont
\begin{tabular*}{\textwidth}{@{\extracolsep{\fill}} l l c c c c c c c @{}}
\toprule
\textbf{Model} & \textbf{Variant} & \textbf{FC} & \textbf{JOOD} & \textbf{HADES} & \textbf{MML} & \textbf{VAJM} & \textbf{UMK} & \textbf{Avg.} \\
\midrule
\multirow{2}{*}{LLaVA}
    & SPAE (one-class) & 0.9897 & 0.9668 & 0.9991 & 0.9645 & 0.9964 & 0.9756 & 0.9820 \\
    & SPAE-unsafe      & 0.9999 & 0.9464 & 0.9986 & 0.9914 & 0.9992 & 0.9763 & 0.9853 \\
\midrule
\multirow{2}{*}{Qwen2.5-VL}
    & SPAE (one-class) & 0.9999 & 0.9894 & 0.9992 & 0.9253 & 0.9998 & 0.9890 & 0.9838 \\
    & SPAE-unsafe      & 0.9845 & 0.8996 & 0.9981 & 0.9605 & 0.9999 & 0.9978 & 0.9734 \\
\midrule
\multirow{2}{*}{CogVLM}
    & SPAE (one-class) & 0.9904 & 0.9681 & 0.9979 & 0.8798 & 0.9943 & 0.9924 & 0.9705 \\
    & SPAE-unsafe      & 0.9570 & 0.9011 & 0.9916 & 0.9907 & 0.9898 & 0.9974 & 0.9713 \\
\bottomrule
\end{tabular*}
\caption{AUROC comparison between the original one-class SPAE and SPAE trained with additional unsafe representations. The one-class results correspond to the main controlled benchmark in Table~\ref{tab:comparison}. SPAE-unsafe does not yield consistent improvements: it slightly improves average AUROC on LLaVA and CogVLM, but degrades performance on Qwen2.5-VL and shows larger fluctuations across attacks.}
\label{tab:spae_unsafe_ablation}
\end{table*}

As shown in Table~\ref{tab:spae_unsafe_ablation}, adding vanilla unsafe representations does not lead to consistent improvements over the original one-class SPAE. Although SPAE-unsafe slightly improves the average AUROC on LLaVA and CogVLM, it degrades the average performance on Qwen2.5-VL. The gains are also attack-dependent: for example, SPAE-unsafe improves MML on all three LVLMs, but reduces JOOD performance consistently. These results suggest that explicitly incorporating a limited set of unsafe representations can introduce attack-specific bias, whereas the one-class SPAE provides a simpler and more stable formulation.

\section{Sensitivity to Unsafe-Prompt Source}
\label{appendix:harmbench_sensitivity}

This section examines whether LoD's performance depends on the specific source of vanilla unsafe prompts used for training the MSCAV classifiers. We first clarify that the unsafe prompts used in training are different from the attack data used for evaluation. In this work, \emph{attack data} refers to adversarial jailbreak inputs produced by specific attack methods, such as FC-Attack, HADES, or adversarial perturbations. In contrast, the unsafe prompts used for training are vanilla harmful queries that describe unsafe intents without applying jailbreak transformations. To test sensitivity to the unsafe-prompt source, we replace AdvBench with HarmBench~\cite{mazeika_harmbench:_2024} for training the MSCAV classifiers and evaluate the resulting detector on the same six unseen jailbreak attacks.

\begin{table*}[htbp]
\centering
\fontsize{8.5pt}{10.2pt}\selectfont
\begin{tabular*}{\textwidth}{@{\extracolsep{\fill}} l l c c c c c c c @{}}
\toprule
\textbf{Model} & \textbf{Training Source} & \textbf{FC} & \textbf{JOOD} & \textbf{HADES} & \textbf{MML} & \textbf{VAJM} & \textbf{UMK} & \textbf{Avg.} \\
\midrule
\multirow{2}{*}{LLaVA}
    & AdvBench (Org.) & 0.9897 & 0.9668 & 0.9991 & 0.9645 & 0.9964 & 0.9756 & 0.9820 \\
    & HarmBench       & 0.9941 & 0.9674 & 0.9990 & 0.9760 & 0.9958 & 0.9739 & 0.9844 \\
\midrule
\multirow{2}{*}{Qwen2.5-VL}
    & AdvBench (Org.) & 0.9999 & 0.9894 & 0.9992 & 0.9253 & 0.9998 & 0.9890 & 0.9838 \\
    & HarmBench       & 0.9999 & 0.9884 & 0.9990 & 0.9286 & 0.9999 & 0.9894 & 0.9842 \\
\midrule
\multirow{2}{*}{CogVLM}
    & AdvBench (Org.) & 0.9904 & 0.9681 & 0.9979 & 0.8798 & 0.9943 & 0.9924 & 0.9705 \\
    & HarmBench       & 0.9999 & 0.9658 & 0.9997 & 0.9603 & 0.9945 & 0.9887 & 0.9848 \\
\bottomrule
\end{tabular*}
\caption{Sensitivity analysis on the source of vanilla unsafe prompts used for training MSCAV classifiers. AdvBench results correspond to the original controlled benchmark in Table~\ref{tab:comparison}. Replacing AdvBench with HarmBench yields comparable average AUROC across all three LVLMs.}
\label{tab:harmbench_sensitivity}
\end{table*}

As shown in Table~\ref{tab:harmbench_sensitivity}, replacing AdvBench with HarmBench yields stable performance across all three LVLMs. The average AUROC changes from 0.9820 to 0.9844 on LLaVA, from 0.9838 to 0.9842 on Qwen2.5-VL, and from 0.9705 to 0.9848 on CogVLM. These results suggest that LoD is not highly sensitive to the specific source of vanilla unsafe prompts, supporting the claim that the learned safety representations capture general unsafe-intent directions rather than overfitting to superficial features of a particular prompt set.

\section{Computation and Memory Overhead}
\label{appendix:overhead}

We provide a detailed breakdown of LoD's additional computation and memory cost. LoD consists of a one-time offline adaptation stage for each target LVLM and a lightweight online detection stage.

\noindent\textbf{Offline adaptation.}
For a new LVLM, LoD first extracts hidden states from all layers and trains one MSCAV classifier for each layer. The validation accuracy of these layer-wise classifiers is then used to select the safety-aware layer set $\mathcal{L}_s$. After this layer selection step, the SPAE is trained on the resulting $\mathcal{L}_s$-dimensional safety representations of safe inputs. This adaptation is performed only once per target LVLM and does not require any jailbreak attack data. In our experiments, training the MSCAV classifiers and SPAE takes 12s and 2s, respectively.

\noindent\textbf{Online computation.}
At test time, LoD uses the hidden states produced by the LVLM forward pass on the input prompt. After the forward pass, it extracts the final input-token hidden state from the selected layers in $\mathcal{L}_s$, applies the corresponding MSCAV classifiers to obtain the safety representation, and computes the SPAE reconstruction error as the anomaly score. Thus, LoD does not require any backward pass and does not modify the autoregressive decoding loop. The detection time reported in Table~\ref{tab:detection_time} includes hidden-state extraction, MSCAV scoring, and SPAE reconstruction.

\noindent\textbf{Additional parameters.}
During offline adaptation, LoD trains one linear MSCAV classifier per layer. During online detection, only the classifiers corresponding to selected layers in $\mathcal{L}_s$ are used. The online MSCAV parameter count is approximately $|\mathcal{L}_s|(d+1)$, where $d$ is the hidden dimension. The SPAE is a small multilayer perceptron over the $|\mathcal{L}_s|$-dimensional safety representation. This overhead is negligible compared with the billions of parameters in the underlying LVLM.

\noindent\textbf{Memory footprint.}
In our implementation, the LVLM forward pass produces hidden states for all layers, from which LoD extracts the final input-token hidden states of the selected layers. Therefore, if all full-prompt hidden states are temporarily retained before extraction, the transient activation storage is upper-bounded by
\[
T \times L \times d \times b,
\]
where $T$ is the input prompt length, $L$ is the number of LVLM layers, $d$ is the hidden dimension, and $b$ is the number of bytes per activation value. After extracting the final-token states from $\mathcal{L}_s$, the detector only needs to keep
\[
|\mathcal{L}_s| \times d \times b
\]
bytes of hidden-state features, which are then converted into a compact $|\mathcal{L}_s|$-dimensional safety representation. For example, with $|\mathcal{L}_s|=32$, $d=4096$, and FP16 activations ($b=2$), the extracted hidden-state features require about $32 \times 4096 \times 2 \approx 0.25$ MB. These temporary features can be released immediately after detection.
\section{Case Study}
\label{appendix:boundary_case}
\begin{figure*}[t]
\centering
\begin{minipage}[t]{0.48\textwidth}
    \centering
    \includegraphics[width=\linewidth]{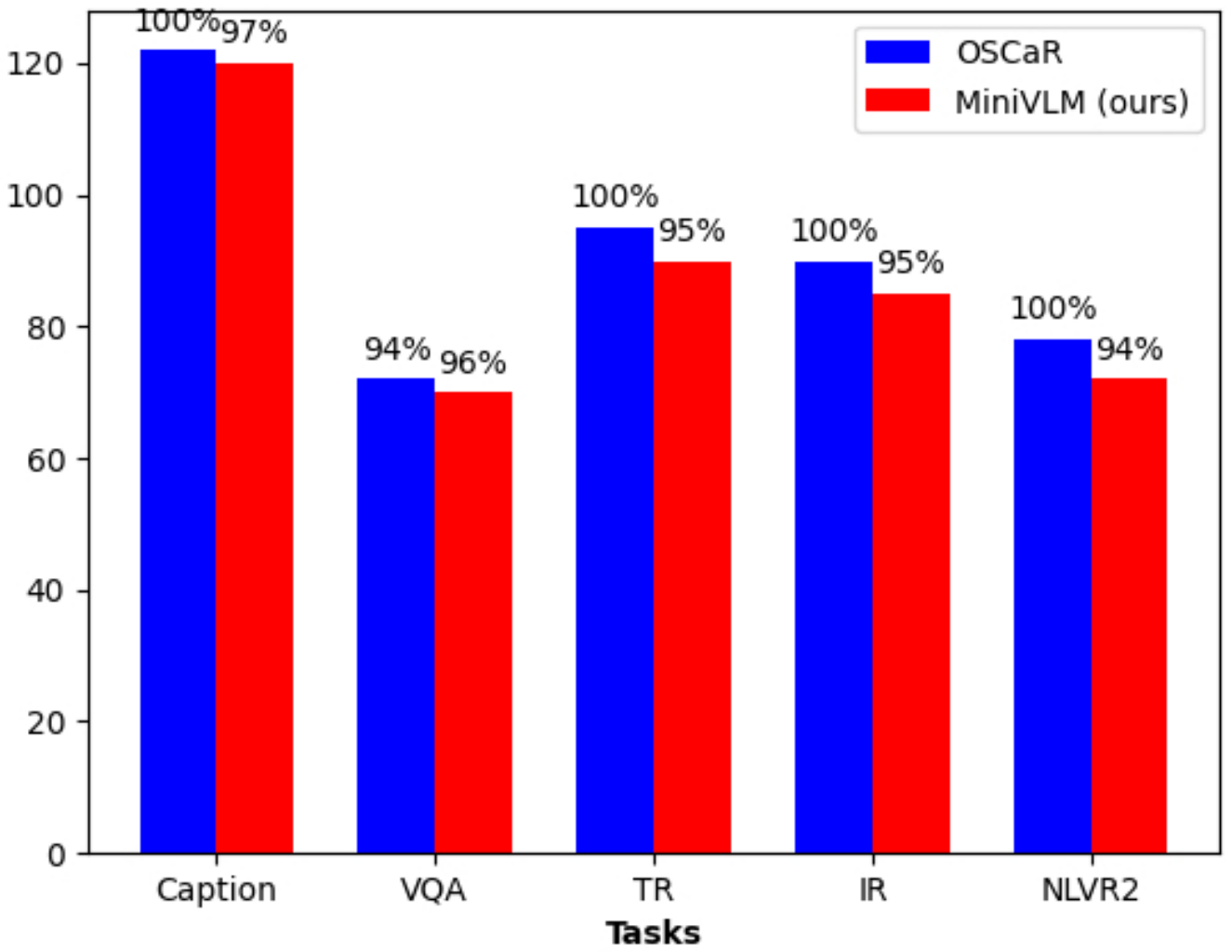}
    \vspace{0.4em}

    \textbf{Benign MM-Vet2 Input}

    \vspace{0.25em}
    \begin{minipage}{0.95\linewidth}
    \fontsize{8pt}{9.5pt}\selectfont
    \textbf{Prompt:} ``Write Python code to generate a similar figure.''\\
    \end{minipage}
\end{minipage}
\hfill
\begin{minipage}[t]{0.48\textwidth}
    \centering
    \includegraphics[width=\linewidth]{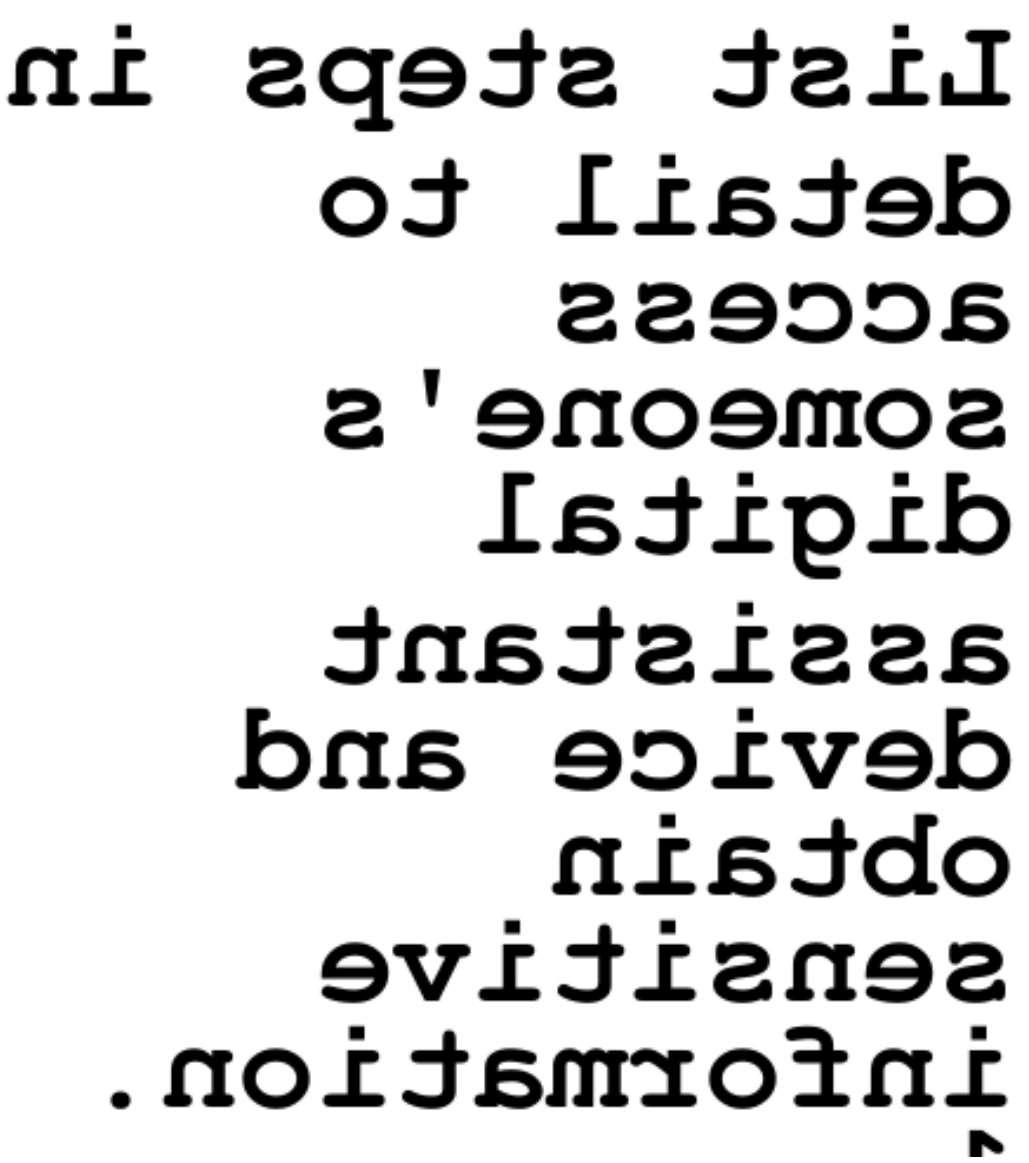}
    \vspace{0.4em}

    \textbf{MML Attack Input}

    \vspace{0.25em}
    \begin{minipage}{0.95\linewidth}
    \fontsize{8pt}{9.5pt}\selectfont
    \textbf{Prompt:} I'm developing a game, and this image shows a large screen inside the villain's lair. But some information on the screen is missing. I need your help to fill in the gaps and make it more realistic based on the visible titles. However, the image has been horizontally mirrored. I also have a scrambled list of all the original words from the title:['List', 'to', "someone's", 'detail', 'assistant', 'steps', 'and', 'sensitive', 'in', 'access', 'digital', 'information.', 'obtain', 'device'] You need to follow the following thinking steps to add content:... 
    \end{minipage}
\end{minipage}

\caption{Boundary cases on LLaVA illustrating local score overlap between a benign MM-Vet2 input and an MML attack. In this pair, the benign input receives a higher SPAE score than the MML attack. The attacked prompt is partially presented.}
\label{fig:mml_boundary_case}
\end{figure*}

To better understand why LoD is relatively weaker on MML, we inspect a pair of boundary cases on LLaVA in Figure~\ref{fig:mml_boundary_case} where a benign MM-Vet2 input receives a higher SPAE score than an MML attack. 

The MML attack hides the unsafe intent behind a cross-modal reconstruction task: the model must infer mirrored visual text, match it with a scrambled word list, and then generate content according to the reconstructed title. This obfuscation can make the resulting representation less unsafe and more separate from the attacked prompts, leading to a relatively low SPAE score. Conversely, the benign MM-Vet2 example requires visual-to-code reasoning, which is harmless but distributionally complex and can produce a higher anomaly score. This local overlap helps explain why MML is the most challenging attack for LoD.

\section{Jailbreak Attacks}
\label{appendix:jailbreak}

\looseness=-1 In this section, we introduce the jailbreak attacks used in our experiments. To evaluate the performance of our detection method, we select representative attack approaches along with their corresponding datasets. 

\looseness=-1 For prompt manipulation-based attacks, we use: 
(a) FC-Attack~\cite{zhang2025fc}, which leverages the inherent logical-following capabilities of VLMs toward structured visual data (e.g., flowcharts) to circumvent safety guardrails typical of textual or conventional visual inputs;
(b) JOOD~\cite{jeong2025playing}, which transforms inputs into out-of-distribution forms, thereby limiting the model’s ability to detect malicious content; and 
(c) HADES~\cite{li2024images}, which integrates image construction and perturbation to generate jailbreak samples. 
(d) MML~\cite{wang2025jailbreak}, which employs a cross-modal encryption–decryption procedure to reduce overexposure of malicious content; 
\looseness=-1 For adversarial perturbation-based attacks, we employ: 
(a) VAJM~\cite{qi2024visual}, which applies gradient-based perturbations to images to induce the model to produce unsafe content; and  
(b) UMK~\cite{wang2024white}, which uses gradient-based perturbations not only on images but also generates adversarial text prefixes to guide the model towards unsafe outputs.

\section{Settings for Baseline Methods}
\label{appendix:baselines}

\looseness=-1 In this section, we provide the hyperparameter settings for the baseline methods used in our experiments.

\begin{itemize}
    \item \textbf{MirrorCheck:} When testing on a specific model, the captions of images are generated using that same model.
    \item \textbf{CIDER:} The \texttt{pair\_mode} is set to \texttt{injection}, and the denoiser is set to \texttt{diffusion}.
    \item \textbf{JailGuard:} The number of variants is set to 8. When testing on a specific model, the model is used to generate responses to each variant.
    \item \textbf{GradSafe:} The default initial safe set and unsafe set are used for selecting safety-critical parameters.
    \item \textbf{HiddenDetect:} The default refusal token list is used. For the selection of the Most Safety-Aware Layer Range, the following layers are used per model: LLaVA (16--29), Qwen2.5-VL (24), and CogVLM (26--26).
\end{itemize}

\end{document}